\documentclass[journal]{IEEEtran}
\usepackage{amsmath, amssymb, cite, tikz, graphicx, xspace, mathrsfs, mathtools, psfrag, chemarrow, subfigure,color,soul,graphicx,mathtools}

\usepackage{kbordermatrix}
\usepackage{enumitem}
\usepackage{bbm}
\usepackage[utf8]{inputenc}
\usepackage{booktabs}
\usepackage{amscd}
\usepackage{amsmath}
\usepackage{amssymb}
\usepackage{amsthm}

\usepackage{epsfig}
\usepackage{verbatim}
\usepackage{amsthm}
\usepackage{color}
\usepackage{ multirow }
\usepackage[noend]{algpseudocode}
\usetikzlibrary{patterns}
\usetikzlibrary{angles} 
\usepackage{eucal}
\usepackage{algorithm}
\usepackage{tikz}
\usetikzlibrary{shapes, arrows, decorations.markings, arrows.meta}
\usetikzlibrary{patterns} 
\usetikzlibrary{arrows,automata,positioning,chains,calc}
\usetikzlibrary{quotes,angles}
\newtheorem{remark}{Remark}
\makeatletter
\def\BState{\State\hskip-\ALG@thistlm}
\makeatother
\newtheorem{theorem}{Theorem}
\newtheorem{lemma}[theorem]{Lemma}
\begin{document}
	\setlength{\abovecaptionskip}{-3pt}
	\setlength{\belowcaptionskip}{1pt}
	\setlength{\floatsep}{1ex}
	\setlength{\textfloatsep}{1ex}
\title{So Timely, Yet So Stale:\\ The Impact of Clock Drift in Real-Time Systems}

\author{Mehrdad Salimnejad, Nikolaos Pappas, Marios Kountouris
\thanks{M. Salimnejad and N. Pappas are with the Department of Computer and Information Science Linköping University, Sweden, email: \{\texttt{mehrdad.salimnejad, nikolaos.pappas\}@liu.se}. M. Kountouris is with the Communication Systems Dept.,
EURECOM, France, and also with the Department of Computer Science and Artificial Intelligence, University of Granada, Spain, email: \texttt{mariosk@ugr.es}. \\ The work of M. Salimnejad and N. Pappas has been supported in part by the Swedish Research Council (VR), ELLIIT, and the European Union (ETHER, 101096526). The work of M. Kountouris has received funding from the European Research Council (ERC) under the European Union’s Horizon 2020 Research and Innovation programme (Grant agreement No. 101003431).
}}

\maketitle

\begin{abstract}
In this paper, we address the problem of timely delivery of status update packets in a real-time communication system, where a transmitter sends status updates generated by a source to a receiver over an unreliable channel. 
The timestamps of transmitted and received packets are measured using separate clocks located at the transmitter and receiver, respectively. To account for possible clock drift between these two clocks, we consider both \emph{deterministic} and \emph{probabilistic} drift scenarios. 
We analyze the system's performance regarding the Age of Information (AoI) and derive closed-form expressions for the distribution and the average AoI under both clock drift models. Additionally, we explore the impact of key system parameters on the average AoI through analytical and numerical results.
 \end{abstract}
 \vspace{-0.3cm}
\section{Introduction}
\par Timely and effective acquisition, transmission, and processing of causal information are critical for distributed systems, autonomous multi-agents systems, and communication networks that support time-sensitive applications, services reliant on status updates, and causal event ordering \cite{kountouris2021semantics,abd2019role,PetarProc2022,popovski2024time,salimnejad2024real}. In these systems, agents continuously observe and transmit update messages from a source to a remote monitor via a communication network. These updates are processed to extract useful (semantically valuable) information, enabling prompt decision-making or actuation. The effectiveness of decision-making in such systems, which directly affects the performance of agentic applications, is highly dependent on the freshness of the information at the destination. The Age of Information (AoI) is a key metric to quantify the timeliness or freshness of information \cite{kosta2017age,sun2019age,yates2021age}. AoI is defined as the time elapsed since the generation of the most recent status update that has been successfully received at the destination. 

To date, the AoI and the timing of events have primarily been measured under the assumption of a common, perfectly synchronized, and error-free timing reference between local clocks.  
In other words, the local clocks at the transmitter and receiver, which are used to timestamp events for transmitting and receiving update packets, are typically assumed to be perfectly synchronized.
However, in real-world systems, achieving precise clock synchronization is challenging due to hardware imperfections, network delays, and other contributing factors \cite{mills1991internet, lamport2019time}. For example, in distributed systems where each node relies on its local clock rather than a global clock, discrepancies between clocks can result in inconsistent timestamps during data exchanges \cite{SUNDARARAMAN2005281}. This clock misalignment can cause delays in the delivery of status updates and reduce the accuracy of AoI measurements, ultimately degrading the performance of real-time systems.
Clock drift and discrepancies can also naturally arise due to time dilation and other relativistic effects, especially in scenarios with significant relative velocity differences between the transmitter and receiver or variations in gravitational potentials \cite{Drake2014}.
These effects are particularly relevant in satellite communications, high-speed networks, space missions, interstellar communication, and high-precision navigation systems, where even minor clock drifts can significantly impact the synchronization and the accuracy of time-sensitive operations \cite{Messerschmitt17}. Research on relativistic effects in communication systems has primarily focused on information transmission efficiency \cite{Kovacevic2024}. However, the impact of relativistic effects on AoI and information freshness is not well understood.

\par This letter explores the impact of misaligned timing references on information aging, with a particular focus on time dilation effects and clock drifts with respect to the AoI. 
Specifically, we analyze a time-slotted communication system with a source, transmitter, and receiver, where the transmitter observes the source and sends status updates over an unreliable channel. Two separate clocks, one at the transmitter and one at the receiver, are used to timestamp transmitted and received packets, respectively. To account for clock drift, we consider both \emph{deterministic} and \emph{probabilistic} drift cases. We evaluate the system performance in terms of AoI and derive expressions for the distribution and the average AoI under both scenarios.

\par This work lays the foundation for introducing the concept of \emph{Referential} or \emph{Relativistic} AoI, where the age and timeliness of information are defined relative to the distinct frames of reference of the transmitter and receiver; these frames may evolve differently due to various factors. Analogous to the relativity and non-universality of time, \emph{referential AoI} captures the relativity and reference-dependence of information freshness. This relativity directly influences the ordering of events and causal decision-making processes. Simply put, just as time is relative to the observer, so too is the timeliness of information.
 \vspace{-0.4cm}
\section{System Model}
\par
\begin{figure}[!h]
	\centering
	\footnotesize
 \includegraphics[width=1.04\linewidth]{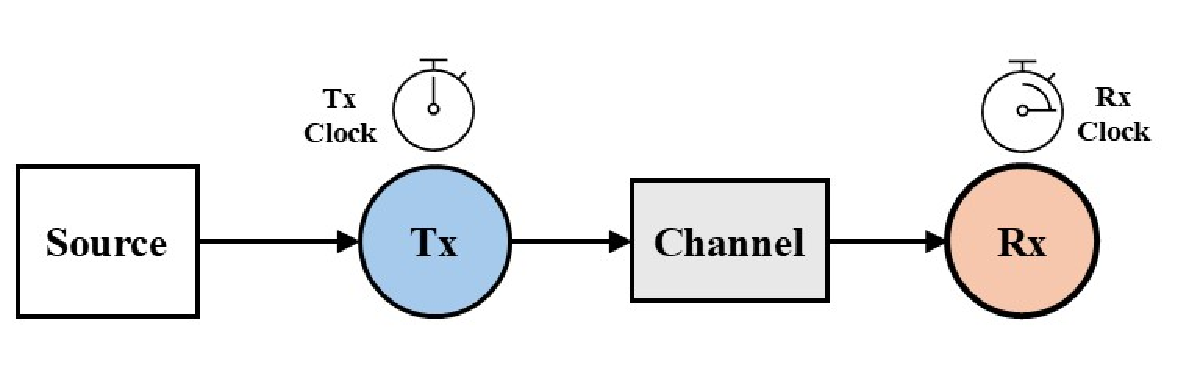}
	\vspace*{2ex}
	\caption{A real-time status update system with local clocks.}
	\label{system_model_fig}
\end{figure}

 We consider a time-slotted communication system where a source generates status updates in the form of packets at each time slot $t$, $\{ t \in \mathbb{N}\}$, as shown in Fig. \ref{system_model_fig}. A transmitter (Tx) sends the generated packets to a remote receiver (Rx) over a wireless communication channel. The channel state at time $t'$ is denoted as $h(t)$, where $h(t) = 1$ represents successful packet decoding by the receiver, and $h(t) = 0$ indicates a decoding failure. The probabilities of successful and failed packet decoding are given by $p_{s} = \mathbb{P}\big[h(t)=1\big]$ and ${p}_{f} = 1-p_{s}$, respectively. 

We assume that there is no single global clock in the system, as is usually the case in distributed systems. Instead, each node maintains its own local clock, meaning there are two distinct clocks: one located at the transmitter and another at the receiver. These clocks are used to measure time and timestamp events, such as when a packet is sent and received. We assume the presence of clock drift (a difference in clock rates) between these two clocks. This drift causes the recorded time at the receiver to deviate from that at the transmitter. Consequently, the receiver's clock may run faster or slower than the transmitter's clock, or both clocks may run differently relative to a reference clock.
In other words, the two clocks are not perfectly synchronized for time measurement, and their readings may differ by some units of time (e.g., milliseconds or seconds).
 \vspace{-0.4cm}
\par We consider both \emph{deterministic} and \emph{probabilistic} scenarios involving clock drift between the transmitter and receiver. In the \emph{deterministic} scenario, we assume a constant clock drift, denoted by $d$ $(\text{where}\hspace{0.1cm} d \in \mathbb{N}_{0})$, in the receiver clock relative to the transmitter clock at every time slot. In the \emph{probabilistic} scenario, we study two distinct cases. In the first case, the receiver experiences only positive drift relative to the transmitter. Specifically, at  time slot $t$, the receiver's clock may drift by $k \in \{0, 1, 2, \ldots, K\}$ slots compared to the transmitter's clock, where $K$ denotes the maximum possible slot drift within the system. We also assume that at each time slot, the slot drift process, denoted by $\delta (t)$, follows a categorical distribution (or generalized Bernoulli) as follows
\begin{align}
    \label{Drift_RV}
    \mathbb{P}[\delta(t) = k] =
    \begin{cases}
    p_{0}, & k =0,\\
    p, & 1\leqslant k \leqslant K
    \end{cases}
\end{align}
where $p_{0}$ and $p$ denote the probabilities that at time slot $t$, the receiver's clock has drifted by $0$ slots and $k \in \{1, 2, \cdots, K\}$ slots relative to the transmitter's clock, respectively, and $p_{0} + Kp = 1$. 
Furthermore, we consider a scenario where the receiver's clock encounters both positive and negative drift relative to the transmitter's clock. Specifically, at time slot $t$, the drift at the receiver can take $k \in \{-1, 0, 1\}$ slots, with a probability $p_k=\mathbb{P}[\delta(t)=k]$, such that $p_{-1} + p_0 + p_1 = 1$.

\section{Performance Metric and Analysis}
\par In this section, we analyze the AoI, a performance metric that measures the time elapsed since the most recent update was generated by the source and successfully received by the receiver. We derive a closed-form expression for the average AoI, considering both deterministic and probabilistic clock drift between the transmitter and receiver clocks.
\subsection{Deterministic Clock Drift}
\par Let $\Delta(t)$ be a positive integer representing the AoI at the receiver at time slot $t$. We define the evolution of AoI where there is a drift of $d \in \mathbb{N}_{0}$ slots between the clocks of the transmitter and the receiver, as follows
\begin{align}
	\label{AoI_fixk}
	\Delta(t) = 
	\begin{cases}
		d+1, &h(t)=1,\\
		\Delta(t-1)+1, &h(t)=0.
	\end{cases}
\end{align}
\begin{lemma}
		\label{theorem_AvgAoI_dfix}
        The average AoI, $\bar{\Delta}$, when the receiver's clock drifts by $d\geqslant 0$ slots relative to the transmitter's clock, is given by
  \begin{align}
      \bar{\Delta} = d+\frac{1}{p_{s}}.
  \end{align}
	\end{lemma}
	
	\begin{IEEEproof} 
		See Appendix \ref{Appendix_Lemma_dfix_AvgAoI}.
	\end{IEEEproof}
\subsection{Probabilistic Clock Drift}
\par In this section, we consider a scenario where, at time slot $t$, the receiver's clock may drift by $k \in \{0, 1, 2, \dots, K\}$ slots relative to the transmitter's clock, with a probability $p_k = \mathbb{P}[\delta(t) = k]$, where $p_k = p$, $\forall 1 \leqslant k \leqslant K$. We can define the evolution of AoI as  
\begin{align}
\label{AoI_Prob_kfix}
	\Delta(t) \!=\! 
	\begin{cases}
		\delta(t)+1, & h(t) = 1,\\
		\max\big\{1,\Delta(t-1)\!+\!\delta(t)\!-\!\delta(t-1)\!+\!1\big\}, &h(t) = 0.
	\end{cases}
\end{align}
Using \eqref{AoI_Prob_kfix}, and applying the total probability theorem, $\mathbb{P}[\Delta(t) = i]$ is calculated as
\begin{align}
    \label{PrDi_probdrift}
    \mathbb{P}[\Delta(t) = i] = \sum_{k=0}^{K}\mathbb{P}[\delta(t) = k, \Delta(t) = i] = \sum_{k=0}^{K} \pi_{k,i},
\end{align}
where $\pi_{k,i}$ is the probability obtained from the stationary distribution of the two-dimensional discrete-time Markov chain (DTMC) describing the joint status of the clock drift regarding the current status of the AoI, i.e., $\big(\delta(t), \Delta(t)\big)$.
\begin{lemma}
		\label{theorem_AvgAoI_prob1}
        The steady state probabilities $\pi_{k,i}$, $\forall k\in\{0,1, \cdots, K\}$ and $i\geqslant 1$ is given by
        \begin{align}
        \label{pik_probdrift}
            \pi_{k,i} =
            \begin{cases}
		p_{0}p_{s}p_{f}^{i-1}, &i\geqslant 1, k =0,\\
		p p_{s}p_{f}^{i-k-1}, & 1\leqslant k \leqslant \min\{K,i-1
        \},\\
        0, &\text{otherwise}.
	\end{cases}
        \end{align}
	\end{lemma}
	
	\begin{IEEEproof} 
		See Appendix \ref{Appendix_Lemma_AvgAoI_prob1}.
	\end{IEEEproof}
Now, using \eqref{pik_probdrift}, we can calculate \eqref{PrDi_probdrift} for $i\geqslant1$ as follows
\begin{align}
    \label{PrDelati_probdrift}
    \mathbb{P}[\Delta(t) \!=\! i] &\!=\! \sum_{k=0}^{K} \pi_{k,i}\notag\\
    &=p_{0}p_{s}p_{f}^{i-1}\!+\!p\Big(\!1\!-\!p_{f}^{\min\{K,i-1\}}\!\Big)p_{f}^{i-1-\min\{K,i-1\}}.
\end{align}
Using \eqref{PrDelati_probdrift}, for the probabilistic clock drift, where the receiver's clock drifts by $k \in \{0, 1, \ldots, K\}$ slots compared to the transmitter's clock, the average AoI, $\bar{\Delta}$, is given by 
\begin{align}
    \label{AvgAoI_lemma}
    \bar{\Delta} = \sum_{i=1}^{\infty} i\mathbb{P}[\Delta(t) = i] = \frac{2+K(K+1)pp_{s}}{2p_{s}}.
\end{align}
\begin{remark}
		\label{remark_Prob1}
		Using \eqref{AvgAoI_lemma} and the condition $p_{0}+Kp = 1$ with $0\leqslant p_{0}\leqslant 1$, we can prove that with the maximum probability $p_{\text{max}}$, the system can tolerate up to $K$ slots of drift while ensuring that $\bar{\Delta}\leqslant \bar{\Delta}_{TH}$, as given by 
  \begin{align}
      \label{probmax}
      p_{\text{max}} = \max\left\{0,\min\left\{\frac{1}{K},\frac{2\big(p_{s}\bar{\Delta}_{TH}-1\big)}{K(K+1)p_{s}}\right\}\right\}.
  \end{align}
	\end{remark}

 \par We now consider a scenario where, at each time slot, the receiver's clock has a drift of $k \in \{-1, 0, 1\}$ slots compared to the transmitter's clock, with probabilities $p_{k} = \mathbb{P}[\delta(t) = k]$. The evolution of AoI is given by
\begin{align}
\label{AoI_p0pmp1}
	\Delta(t) \!=\! 
	\begin{cases}
		\max\{1,\delta(t)+1\}, & \!\!h(t) = 1,\\
		\max\big\{1,\Delta(t-1)\!+\!\delta(t)\!-\!\delta(t-1)\!+\!1\big\}, &\!\!h(t) = 0.
	\end{cases}
\end{align}
Using \eqref{AoI_p0pmp1}, and applying the total probability theorem, we can express the probability that the AoI at time slot $t$ equals $i\geqslant 1$ as follows
\begin{align}
    \label{PrAoI_i_prob2}
    \mathbb{P}[\Delta(t) \!=\! i] &\!=\! \mathbb{P}[\delta(t) \!=\! -1, \Delta(t)\! =\! i]\!+\!\mathbb{P}[\delta(t) \!=\! 0, \Delta(t) \!=\! i]\notag\\
    &+\mathbb{P}[\delta(t)\!=\!1,\Delta(t) \!=\! i]=\pi_{-1,i}+\pi_{0,i}+\pi_{1,i}.
\end{align}
Note that $\pi_{-1,i}$, $\pi_{0,i}$, and $\pi_{1,i}$ in \eqref{PrAoI_i_prob2} represent probabilities derived from the stationary distribution of the two-dimensional DTMC describing the joint status of the slot drift regarding the current state of the AoI, i.e., $\big(\delta(t), \Delta(t)\big)$.
\begin{lemma}
		\label{theorem_AvgAoI_prob2}
        The steady state probabilities $\pi_{k,i}$, $\forall k\in\{-1,0,1\}$ and $i\geqslant 1$, are given by
\begin{align}
	\label{AvgAoI_p0pmp1_1_lemma}
\pi_{k,i} \!=\!
\begin{cases}
 p_{-1}p_{s}, &\!\!\!\!k=-1, i=1,\\
    p_{-1}p_{s}p_{f}^{i-1}F, &\!\!\!\!k=-1, i\geqslant 2,\\
    p_{k} p_{s}, &\!\!\!\!k\!\in\!\{0,1\}, i=k+1,\\
    p_{k}p_{s}p_{f}(1-p_{-1}), &\!\!\!\!k\!\in\!\{0,1\}, i=k+2,\\
    p_{k}p_{s}p_{f}^{i-2-k}F,&\!\!\!\! k\!\in\!\{0,1\}, i\geqslant k+3.
\end{cases}
	\end{align}
    where $F$ in \eqref{AvgAoI_p0pmp1_1_lemma} is given by
    \begin{align}
        \label{Fx}
        F = 1-p_{s}(1-p_{-1}).
    \end{align}
	\end{lemma}
	\begin{IEEEproof} 
		See Appendix \ref{Appendix_Lemma_AvgAoI_prob2}.
	\end{IEEEproof}
Now, using \eqref{AvgAoI_p0pmp1_1_lemma}, \eqref{PrAoI_i_prob2} can be written as
\begin{align}
    \label{ProbAoI_i_prob2_final}
    \mathbb{P}[\Delta(t) \!=\! i]\!=\!\!
    \begin{cases}
       \! p_{0}p_{s}+p_{-1}p_{s}\big(1+p_{f}\big(1-p_{-1}\big)\big),&\! \!\!\!\!i \!=\!1,\\
        \!p_{-1}p_{s}p_{f}F\!+\!p_{0}p_{s}p_{f}(1\!-\!p_{-1})\!+\!p_{1}p_{s},&\!\!\!\!\!i\!=\!2,\\
        \!\!\big(p_{-1}p_{s}p_{f}^{2}\!+\!p_{0}p_{s}p_{f}\big)F\!+\!p_{1}p_{s}p_{f}(1\!-\!p_{-1}),&\!\!\!\!\!i\!=\!3,\\
        \!\!\big(p_{-1}p_{s}p_{f}^{i-1}\!+\!p_{0}p_{s}p_{f}^{i-2}\!+\!p_{1}p_{s}p_{f}^{i-3}\big)F,&\!\!\!\!\!i\!\geqslant\! 4.
    \end{cases}
\end{align}
where $F$ in \eqref{ProbAoI_i_prob2_final} is obtained in \eqref{Fx}.
\par Using \eqref{ProbAoI_i_prob2_final}, for the probabilistic clock drift, where the receiver's clock drifts by $-1, 0$, and $1$ slots with probabilities $p_{-1}, p_{0}$, and $p_{1}$ compared to the transmitter's clock, the average AoI, $\bar{\Delta}$, is given by
\begin{align}
    \label{AvgAoI_Prob2_WMinus}
    \bar{\Delta} &= \sum_{i=1}^{\infty} i \mathbb{P}[\Delta(t)=i]= p_{1}+\frac{1}{p_{s}}.
\end{align}
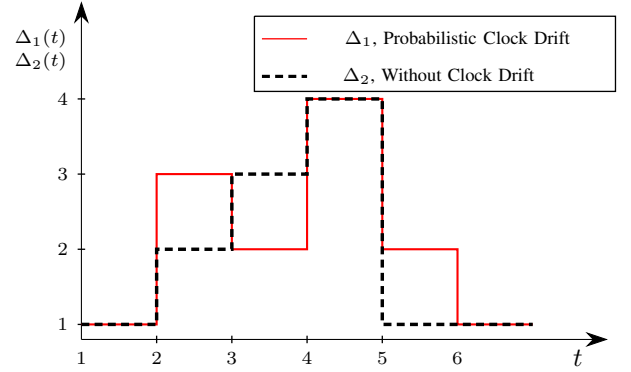
\begin{figure}
  	\centering 
  	
  	\begin{tikzpicture}[scale=1]
  		\draw [-{Stealth[length=3mm, width=2mm]}](0,0) -- (7,0);
  		\draw [-{Stealth[length=3mm, width=2mm]}](0,0) -- (0,4.5);
  		\node(a) at (6.6,-0.25)  {$t$};
  		\node(a1) at (0,-0.25)  {\scriptsize${1}$};
  		\node(a1) at (1,-0.25)  {\scriptsize${2}$};
  		\draw[line width=0.5pt,black] (1,-0.05)--(1,0.05);
  		\node(a1) at (2,-0.25)  {\scriptsize${3}$};
    \draw[line width=0.5pt,black] (2,-0.05)--(2,0.05);
  		\node(a1) at (3,-0.25)  {\scriptsize${4}$};
  		\draw[line width=0.5pt,black] (3,-0.05)--(3,0.05);
  		\node(a1) at (4,-0.25)  {\scriptsize${5}$};
    \draw[line width=0.5pt,black] (4,-0.05)--(4,0.05);
  		\node(a1) at (5,-0.25)  {\scriptsize${6}$};
  		\node(b1) at (-0.25,0.2)  {\scriptsize${1}$};
  		\draw[line width=0.5pt,black] (-0.05,0.2)--(0.05,0.2);
  		\node(b1) at (-0.25,1.2)  {\scriptsize${2}$};
  		\draw[line width=0.5pt,black] (-0.05,1.2)--(0.05,1.2);
    
  		\node(b1) at (-0.25,2.2)  {\scriptsize${3}$};
  		\draw[line width=0.5pt,black] (-0.05,2.2)--(0.05,2.2);
  		\node(b1) at (-0.25,3.2)  {\scriptsize${4}$};
            \draw[line width=0.5pt,black] (-0.05,3.2)--(0.05,3.2);
           
  		\draw[line width=0.8pt, red] (0,0.2) -- (1,0.2) -- (1,0.2)--(1,2.2)--(2,2.2)--(2,1.2)--(3,1.2)--(3,3.2)--(4,3.2)--(4,1.2)--(5,1.2)--(5,0.2)--(6,0.2);
  		\draw[densely dashed, line width=1.3pt] (0,0.2) -- (1,0.2)--(1,1.2)--(2,1.2)--(2,2.2)--(3,2.2)--(3,3.2)--(4,3.2)--(4,0.2)--(5,0.2)--(6,0.2);
  		\draw[line width=0.5pt, red] (2.4,4)--(3,4);
  		\node(l1) at (5,4)  {\scriptsize{$\Delta_{1}$, Probabilistic Clock Drift}};
  		\draw[densely dashed, line width=1.2pt] (2.4,3.5)--(3.,3.5);
  		\node(l1) at (4.75,3.5)  {\scriptsize{$\Delta_{2}$, Without Clock Drift}};
            \draw[line width=0.5pt, black] (2.3,4.3)--(7.1,4.3)--(7.1,3.3)--(2.3,3.3)--(2.3,4.3);
  		\node(y1) at (-0.54,4.)  {\scriptsize{$\Delta_{1}$}$(t)$};
    \node(y2) at (-0.54,3.7)  {\scriptsize{$\Delta_{2}$}$(t)$};
  	\end{tikzpicture}
  	\caption{The evolution of the AoI metric considering probabilistic clock drift and the absence of clock drift.}
   \label{FigCompare}
  \end{figure}
\par To clarify the evolution of the AoI metric with and without clock drift, Fig. \ref{FigCompare} presents an example illustrating the evolution of this metric under probabilistic clock drift where the receiver's clock has drifted to $-1$, $0$, and $1$ slots. In this figure, we assume that $\Delta_{1}(1) = 1$ and $\Delta_{2}(1) = 1$, where $\Delta_{1}(t)$ and $\Delta_{2}(t)$ represent the AoI at time slot $t$ considering the probabilistic clock drift and without clock drift, respectively. At $t = 2$, we have a failed transmission with a $1$ slot drift. Therefore, $\Delta_{1}(2) = 3$ and $\Delta_{2}(2) = 2$. At $t = 3$, we have a failed transmission and a $-1$ slot drift, thus $\Delta_{1}(3) = 2$ and $\Delta_{2}(3) = 3$. At $t = 4$, we do not have a successful transmission and there is a $0$ slot drift; therefore, $\Delta_{1}(4) = 4$ and $\Delta_{2}(4) = 4$. At $t = 5$, we have a successful transmission with a $1$ slot drift; therefore, $\Delta_{1}(5) = 2$ and $\Delta_{2}(5) = 1$. At $t = 6$, we have a successful transmission and there is no slot drift; therefore, $\Delta_{1}(6) = 1$ and $\Delta_{2}(6) = 1$.

\begin{remark}
The above analysis can also be performed using the referential or relativistic AoI (rAoI), defined as follows: let observer $P'$ (e.g., the receiver) be located in the reference frame (coordinate system) $(x',y',z',t')$, while observer $P$ is located in reference frame $(x,y,z,t)$; frames can be inertial or non-inertial with the common assumption that at $t=t'=0$ the two systems coincide. In this context, AoI becomes reference-dependent and may be measured differently depending on the chosen frame of reference. The AoI perceived at the receiver is given by $\Delta(t) = \max\{1, t' - u(t)\}$ where $u(t)$ is the time when the last update was generated in the transmitter’s frame, and $t'$ is the time it was successfully received at the receiver. By expressing the coordinates in vector form, the relationship between the two systems can be described through a transformation $M'=\Lambda M$, where $\Lambda$ is a matrix that can depend on various system parameters and system dynamics. For instance, in special relativity (flat space-time), the Lorentz transformation can be applied to relate different rAoIs. In the analysis above, we considered a non-relativistic scenario where $t' = t + s$, with $s$ representing the slot drift between the transmitter's and receiver's clocks.
\end{remark}
\section{Numerical Results}
\par In this section, we validate our analytical results and assess the performance of the average AoI under probabilistic clock drift. Simulation results are obtained by averaging over $10^6$ time slots, with the initial value of the AoI set to $\Delta(1)=1$.
\begin{figure}[!t]
	\centering
	\includegraphics[width=1.04\linewidth]{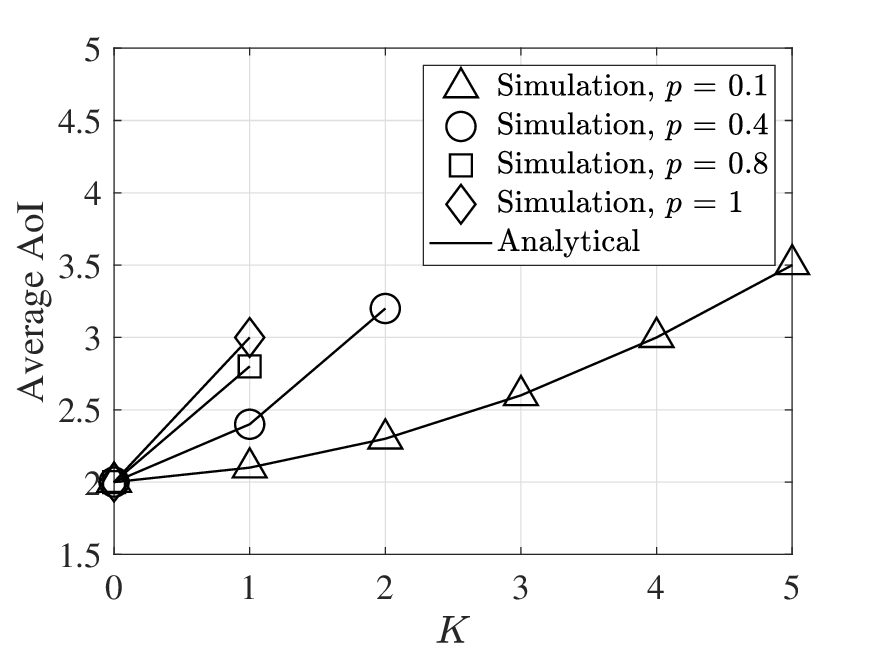}
	\caption{Average AoI as a function of $K$ for $p_{s} = 0.5 $ and $p = 0.1, 0.4, 0.8, 1$.}
	\label{AvgAoIProb}
\end{figure}

\par Fig. \ref{AvgAoIProb} shows the average AoI as a function of $K$ for $p_{s} = 0.5$ and different values of $p$. As seen in this figure, for each value of $p$, when $K$ increases, the average AoI increases. This is because $K$ is the maximum number of possible slot drifts in the system. Therefore, when $K$ increases, with the probability of $p$, the receiver's clock has $k \in \{1, 2, \cdots, K\}$ slots drift compared to the transmitter's clock, which increases the average AoI. Furthermore, as $p$ increases, the average AoI does not have a value for high values of $K$. The reason is that the values of $K$ and $p$ must satisfy $p_{0} = 1 - Kp$ where $0  \leqslant p_{0}  \leqslant 1$. High values of $p$ and $K$ result in $p_{0} < 0$; therefore, it is not feasible for the system to have a large slot drift with high probability. In practice, $K$ represents the maximum allowable synchronization error or clock drift between the transmitter and receiver. A larger
$K$ provides greater flexibility in handling clock drift but can lead to a higher average AoI, potentially degrading performance in time-sensitive systems. Similarly, $p$ represents the probability of clock drift, where a higher $p$ indicates a greater likelihood of frequent synchronization errors. Understanding the trade-off between $K$, $p$, and the average AoI is crucial for designing systems that effectively balance synchronization requirements and overall performance, particularly in applications requiring precise timing.
\begin{figure}[!t]
	\centering
	\includegraphics[width=1.04\linewidth]{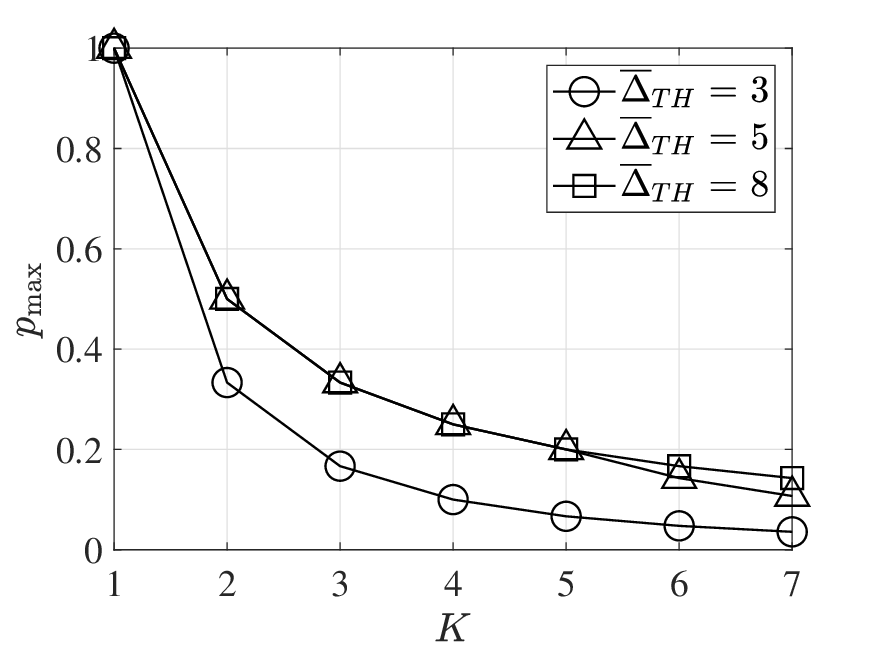}
	\caption{$p_{\text{max}}$ as a function of $K$ for $p_{s} = 0.5 $ and $\bar{\Delta}_{TH} = 3, 5, 8$.}
	\label{pmax}
\end{figure}
Fig. \ref{pmax} illustrates the maximum probability, $p_{\text{max}}$, given in \eqref{probmax} as a function of $K$ for $p_{s} = 0.5$ and selected values of $\bar{\Delta}_{TH}$. We observe that as $K$ increases, the maximum probability that the system can tolerate $K$ slots of drift in order to have $\bar{\Delta} \leqslant \bar{\Delta}_{TH}$ decreases. This is because increasing $K$ leads to a higher average AoI, which in turn reduces the probability that the average AoI falls below a threshold $\bar{\Delta}_{TH}$.
\section{Concluding Remarks}
\par 
We investigated a time-slotted communication system where a transmitter forwards status updates, generated by a source, in packet form over an unreliable wireless channel to a receiver. We considered the presence of clock drifts between the transmission and reception timestamps of status updates. We analyzed two types of clock drift: deterministic and probabilistic, and assessed their impact on the average AoI. Our findings demonstrated that while clock drift degrades the average AoI, it is possible to determine the probability that the system can tolerate up to $K$ slots of drift while maintaining the average AoI below a specified threshold. These results are particularly relevant in scenarios involving asynchronous clocks, the absence of a common (and perfectly synchronized clock) between the two ends, systems with low-cost hardware (e.g., crystal-free IoT devices), and more intriguingly, \textit{scenarios affected by relativistic time dilation caused by differences in relative speed or gravitational potential between the transmitter and the receiver.} 
Understanding the impact of clock drifts on information freshness is essential for designing reliable distributed systems in time-sensitive applications, including remote monitoring and status update services, satellite navigation, and decision-making with relativistic observers.
\vspace{-0.3cm}
\appendix
	\subsection{Proof of Lemma {\ref{theorem_AvgAoI_dfix}}}
	\label{Appendix_Lemma_dfix_AvgAoI}
 Using the total probability theorem, we can express the steady state probability $\pi_{i}$, which represents the probability that the AoI at time slot $t$ equals $i \geqslant 1$, as follows
 
\begin{align}
	\label{Pi_kfix}
	\pi_{i} &\!=\! \mathbb{P}[\Delta(t)=i]\notag\\
	&\!=\!\sum_{j=1}^{\infty}\mathbb{P}\big[\Delta(t)\!=\!i\big|\Delta(t\!-\!1)\!=\!j\big]\mathbb{P}\big[\Delta(t\!-\!1)\!=\!j\big].
\end{align}
When $d\geqslant 0$, using \eqref{AoI_fixk}, the expression given in \eqref{Pi_kfix} can be obtained as
\begin{align}
	\label{Pi_kfix2}
	\pi_{i} = 
	\begin{cases}
		p_{s}({p}_{f})^{i-(d+1)},&i\geqslant d+1,\\
		0,&1\leqslant i\leqslant d.
	\end{cases}
\end{align}
 Now, using \eqref{Pi_kfix2}, the average AoI, $\bar{\Delta}$, is given by
\begin{align}
	\bar{\Delta} = \sum_{i=1}^{\infty} i \pi_{i} = d+\frac{1}{p_{s}}.
\end{align}

\subsection{Proof of Lemma {\ref{theorem_AvgAoI_prob1}}}
	\label{Appendix_Lemma_AvgAoI_prob1}
To obtain $\pi_{k,i}$, we depict the two-dimensional DTMC describing the joint status of the slot drift regarding the
current state of the AoI, i.e., $\big(\delta(t), \Delta(t)\big)$ in Fig. \ref{AoI_prob_k_DTMC}, where the transition probabilities $P_{i,j/m,n}=\mathbb{P}[\delta(t)=m, \Delta(t) = n|\delta(t)=i,\Delta(t)=j]$, $\forall m,i \in\{0, 1, 2, \cdots, K\}$ and $j, n\in\{1,2,\cdots\}$ are given by
\begin{align}
    \label{transprob_k}
    P_{i,j/m,n}= 
    \begin{cases}
        p_{0}p_{s},& m=0, i=1,\\
        pp_{s}, & m\neq 0, n=m+1,\\
        p_{0}p_{f},& m=0,n=j+1-i,\\
        pp_{f},& m\neq 0, n=j+1+m-i,\\
        0, & \text{otherwise}.
    \end{cases}
\end{align}
Now, the stationary distribution can be obtained by solving \emph{balance equations} as follows
\begin{align}
    \label{BalanceEq}
    \pi P_{I} = \pi, \hspace{0.2cm}\sum_{k=0}^{K}\sum_{i=k+1}^{\infty}\pi_{k,i} = 1,
\end{align}
where $P_{I}$ in \eqref{BalanceEq} represents the transition probability matrix, with its elements defined by \eqref{transprob_k}. Additionally, $\pi$ denotes the stationary distribution, expressed as the row vector $\pi = [\pi_{0,1}, \pi_{0,2},\cdots, \pi_{1,2}, \pi_{1,3},\cdots, \pi_{K,K+1},\pi_{K,K+2},\cdots]$.
Now, using \eqref{transprob_k} and \eqref{BalanceEq}, we can derive the steady state probabilities $\pi_{k,i}$ $\forall k\in\{0,1,\cdots,K\}$ and $i\geqslant 1$ as follows
\begin{align}
	\label{AvgAoI_p0pmp1_1_lemma_proof}
            \pi_{k,i} =
            \begin{cases}
		p_{0}p_{s}p_{f}^{i-1}, &i\geqslant 1, k =0,\\
		p p_{s}p_{f}^{i-k-1}, & 1\leqslant k \leqslant \min\{K,i-1
        \},\\
        0, &\text{otherwise}.
	\end{cases}
        \end{align}
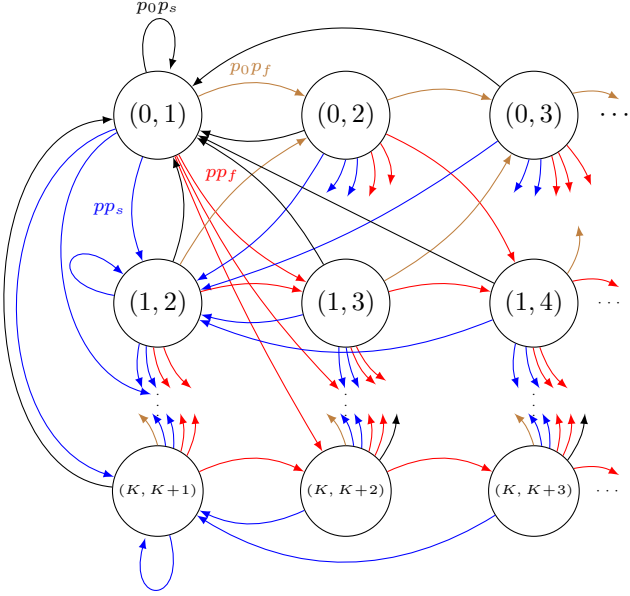
\begin{figure}[!t]
	\centering
	\begin{tikzpicture}[start chain=going left,->,>=latex,node distance=2.2cm,on grid,auto]
	
\node[on chain]             at (3.9,0)(h1) {$\cdots$};
\node[state, on chain]   at (5,0) (3) {$(0,3)$};
			\node[state, on chain]              at (2.5,0) (2) {$(0,2)$};
			\node[state, on chain]             at (0,0)(1) {$(0,1)$};
			\node[state, on chain]             at (0,-2.5)(4) {$(1,2)$};
			\node[state, on chain]             at (2.5,-2.5)(5) {$(1,3)$};
            \node[state, on chain]             at (5,-2.5)(6) {$(1,4)$};
            \tiny{
            \node[state, on chain]             at (0,-5)(7) {$(K,K\!+\!1)$};
            \node[state, on chain]             at (2.5,-5)(8) {$(K,K\!+\!2)$};
             \node[state, on chain]             at (5,-5)(9) {$\tiny(K,K\!+\!3)$};}
            
			\node[on chain]             at (6,-2.5)(h2) {{$\bold{\cdots}$}};
            \node[on chain] at (6.2,-4.8)(h2p) {};
            \node[on chain] at (6.2,-2.3)(h3p) {};
            \node[on chain] at (5.6,-1.4)(h4p) {};
            \node[on chain]             at (0,-3.75)(h3) {$\vdots$};
            \node[on chain]             at (2.5,-3.75)(h4) {$\vdots$};
            \node[on chain]             at (5,-3.75)(h5) {$\vdots$};
            \node[on chain] at (2.2,-1.1)(b1) {};
            \node[on chain] at (2.5,-1.15)(b2) {};
            \node[on chain] at (3.1,-1)(b3) {};
            \node[on chain] at (2.8,-1.15)(b4) {};
            \node[on chain] at (4.7,-1.1)(c1) {};
            \node[on chain] at (5,-1.15)(c2) {};
            \node[on chain] at (5.3,-1.15)(c3) {};
            \node[on chain] at (5.5,-1.1)(c4) {};
            \node[on chain] at (5.8,-1)(c5) {};
            \node[on chain] at (6.2,0.2)(f3) {};
            \node[on chain] at (0,-3.7)(g1) {};
             \node[on chain] at (0.5,-3.7)(g2) {};
             \node[on chain] at (2.7,-3.7)(g3) {};
             \node[on chain] at (3.1,-3.6)(g4) {};
             \node[on chain] at (5.1,-3.7)(g5) {};
             \node[on chain] at (5.5,-3.7)(g6) {};
             \node[on chain] at (-0.2,-3.7)(d1) {};
            \node[on chain] at (0.2,-3.7)(d2) {};
            \node[on chain] at (2.5,-3.7)(e1) {};
            \node[on chain] at (2.95,-3.65)(e2) {};
            \node[on chain] at (4.8,-3.7)(e3) {};
            \node[on chain] at (5.3,-3.7)(e4) {};
            \node[on chain] at (-0.3,-3.9)(k1) {};
            \node[on chain] at (0.3,-3.9)(k2) {};
            \node[on chain] at (0.5,-3.9)(k3) {};
            \node[on chain] at (0.1,-3.9)(k4) {};
            \node[on chain] at (-0.1,-3.9)(k5) {};
            \node[on chain] at (2.2,-3.9)(k6) {};
            \node[on chain] at (2.8,-3.9)(k7) {};
            \node[on chain] at (3,-3.9)(k8) {};
            \node[on chain] at (2.6,-3.9)(k9) {};
            \node[on chain] at (2.4,-3.9)(k10) {};
             \node[on chain] at (3.2,-3.9)(k11) {};
             \node[on chain] at (4.7,-3.9)(k12) {};
            \node[on chain] at (5.3,-3.9)(k13) {};
            \node[on chain] at (5.5,-3.9)(k14) {};
            \node[on chain] at (5.1,-3.9)(k15) {};
            \node[on chain] at (4.9,-3.9)(k16) {};
             \node[on chain] at (5.7,-3.9)(k17) {};
            \tiny	
		\node[on chain]             at (6,-5)(h2) {$\cdots$};
        \scriptsize
		\draw[>=latex]
		(1)   edge[loop above] node {$p_{0}p_{s}$}   (1)
		(1) edge  [bend left=25, brown] node {$p_{0}p_{f}$} (2)
		(1) edge  [bend right=20,left, blue] node {$pp_{s}$} (4)
        (1) edge  [bend right=17,red] node [pos=0.2] {$pp_{f}$} (5)
     (1) edge  [bend right=72,left, blue] node {} (7)
     (1) edge  [bend right=4,left, red] node {} (8)
       (1) edge  [bend right=67,left,blue] node [pos=0.1,left]{} (h3)
      (1) edge  [bend right=9,left, red] node {} (h4)
       (2) edge  [bend left=20,above] node {}(1)
       (2) edge  [bend left=20,brown] node {}(3)
       (2) edge  [bend left=15,blue] node [pos=0.1,left]{}(4)
       (2) edge  [bend left=15,blue] node [pos=0.1,left]{}(b1)
        (2) edge  [bend left=15,blue] node [pos=0.1,left]{}(b2)
        (2) edge  [bend left=20,right,red] node {}(6)
        (2) edge  [bend left=20,right,red] node {}(b3)
        (2) edge  [bend left=20,right,red] node {}(b4)
        (3) edge  [bend right=40,above] node {}(1)
        (3) edge  [bend left=9,blue] node [pos=0.1,left]{}(4)
        (3) edge  [bend left=9,blue] node [pos=0.1,left]{}(c1)
        (3) edge  [bend left=9,blue] node [pos=0.1,left]{}(c2)
        (3) edge  [bend left=9,red] node [pos=0.1,left]{}(c3)
        (3) edge  [bend left=9,red] node [pos=0.1,left]{}(c4)
        (3) edge  [bend left=9,red] node [pos=0.1,left]{}(c5)
        (3) edge  [bend left=20,brown] node {}(f3)
        (4) edge  [bend right=20,above] node {}(1)
        (4) edge  [bend left=15,above,red] node {}(5)
        (4) edge  [loop,out=170, in=140, looseness=8,blue] node {}(4)
        (4) edge  [bend right=15,right, blue] node [pos=0.1]{}(g1)
        (4) edge  [bend right=15,right, blue] node [pos=0.1]{}(d1)
        (4) edge  [bend right=15,right, red] node [pos=0.1]{}(g2)
        (4) edge  [bend right=15,right, red] node [pos=0.1]{}(d2)
         (4) edge  [bend left=15,brown] node [pos=0.1,left]{}(2)
        (5) edge  [bend right=15,above] node {}(1)
        (5) edge  [bend left=15, blue] node [pos=0.1]{}(4)
        (5) edge  [bend right=9,right, blue] node [pos=0.1]{}(g3)
         (5) edge  [bend right=9,right, blue] node [pos=0.1]{}(e1)
        (5) edge  [bend left=15, red] node [pos=0.1]{}(6)
        (5) edge  [bend right=15,right, red] node [pos=0.1]{}(g4)
        (5) edge  [bend right=15,right, red] node [pos=0.1]{}(e2)
        (5) edge  [bend right=15,brown] node [pos=0.1,left]{}(3)
        (6) edge  [right=15,above] node {}(1)
        (6) edge  [bend left=22, blue] node [pos=0.1]{}(4)
        (6) edge  [bend right=15, blue] node [pos=0.1]{}(g5)
        (6) edge  [bend right=15, red] node [pos=0.1]{}(g6)
        (6) edge  [bend right=15, blue] node [pos=0.1]{}(e3)
        (6) edge  [bend right=15, red] node [pos=0.1]{}(e4)
        (6) edge  [bend left=20, red] node [pos=0.1]{}(h3p)
        (6) edge  [bend right=20, brown] node [pos=0.1]{}(h4p)
        (7) edge  [bend left=25, red] node [pos=0.1]{}(8)
        (7) edge  [bend left=85, black] node [pos=0.1]{}(1)
         (7) edge  [loop below,blue] node {}(7)
         (7) edge  [bend right=15, brown] node [pos=0.1]{}(k1)
         (7) edge  [bend right=15, red] node [pos=0.1]{}(k2)
         (7) edge  [bend right=15, red] node [pos=0.1]{}(k3)
         (7) edge  [bend right=15, blue] node [pos=0.1]{}(k4)
         (7) edge  [bend right=15, blue] node [pos=0.1]{}(k5)
         (8) edge  [bend left=25, blue] node [pos=0.1]{}(7)
         (8) edge  [bend left=25, red] node [pos=0.1]{}(9)
         (8) edge  [bend right=15, brown] node [pos=0.1]{}(k6)
         (8) edge  [bend right=15, red] node [pos=0.1]{}(k7)
         (8) edge  [bend right=15, red] node [pos=0.1]{}(k8)
         (8) edge  [bend right=15, blue] node [pos=0.1]{}(k9)
         (8) edge  [bend right=15, blue] node [pos=0.1]{}(k10)
         (8) edge  [bend right=15, black] node [pos=0.1]{}(k11)
         (9) edge  [bend right=15, brown] node [pos=0.1]{}(k12)
         (9) edge  [bend right=15, red] node [pos=0.1]{}(k13)
         (9) edge  [bend right=15, red] node [pos=0.1]{}(k14)
         (9) edge  [bend right=15, blue] node [pos=0.1]{}(k15)
         (9) edge  [bend right=15, blue] node [pos=0.1]{}(k16)
         (9) edge  [bend right=15, black] node [pos=0.1]{}(k17)
         (9) edge  [bend left=20, red] node [pos=0.1]{}(h2p)
         (9) edge  [bend left=32, blue] node [pos=0.1]{}(7)
		;
	\end{tikzpicture}
	\caption{Two-dimensional DTMC describing the joint status of the slot drift regarding the state of the AoI for the probabilistic clock drift, where the receiver's clock at each time slot, with probability $p$, has a $k$-slot drift with $k \geqslant 0$ compared to the transmitter's clock. Arrows with the same color represent transitions with equal probabilities.}
	\label{AoI_prob_k_DTMC}
\end{figure}
\subsection{Proof of Lemma {\ref{theorem_AvgAoI_prob2}}}
	\label{Appendix_Lemma_AvgAoI_prob2}
For the probabilistic clock drift, where at each time slot the receiver's clock experiences a drift of $k \in \{-1, 0, 1\}$ slots relative to the transmitter's clock, the steady state probabilities $\pi_{k,i}$ $\forall i\geqslant 1$ can be derived similarly to Lemma \ref{theorem_AvgAoI_prob1}. In this scenario, the transition probabilities $P_{i,j/m,n} = \mathbb{P}[\delta(t+1) = m, \Delta(t+1) = n \mid \delta(t) = i, \Delta(t) = j]$, $\forall m,i \in\{-1, 0, 1\}$ and $j, n\in\{1,2,\cdots\}$ are given as follows:
\begin{align}
        P_{i,j/m,n} = 
        \begin{cases}
            p_{-1},& m=-1, n=1, j =i+1,\\
            p_{-1}p_{s}& m=-1, n=1, j\geqslant i+2,\\
            p_{-1}p_{f},& m=-1, n=j-i,\\
            p_{0}p_{s},&n=1,\\
            p_{0}p_{f},&m=0, n= j+1-i,\\
            p_{1}p_{s},&m=1, n=2,\\
            p_{1}p_{f},& n=j+2-i.
        \end{cases}
    \end{align}
    Now, using the balance equations in \eqref{BalanceEq}, the steady state probabilities $\pi_{k,i}$ $\forall k\in\{-1,0,1\}$ and $i\geqslant 1$ can be calculated as
    \begin{align}
	\label{AvgAoI_p0pmp1_1_lemma_proof}
\pi_{k,i} \!=\!
\begin{cases}
 p_{-1}p_{s}, &\!\!\!\!k=-1, i=1,\\
    p_{-1}p_{s}p_{f}^{i-1}F, &\!\!\!\!k=-1, i\geqslant 2,\\
    p_{k} p_{s}, &\!\!\!\!k\!\in\!\{0,1\}, i=k+1,\\
    p_{k}p_{s}p_{f}(1-p_{-1}), &\!\!\!\!k\!\in\!\{0,1\}, i=k+2,\\
    p_{k}p_{s}p_{f}^{i-2-k}F,&\!\!\!\! k\!\in\!\{0,1\}, i\geqslant k+3.
\end{cases}
	\end{align}
    where $F=1-p_{s}(1-p_{-1})$.
\bibliographystyle{IEEEtran}
\bibliography{ref}
\end{document}